\documentclass[sigconf]{acmart}
\renewcommand\footnotetextcopyrightpermission[1]{} 

\AtBeginDocument{%
  \providecommand\BibTeX{{%
    \normalfont B\kern-0.5em{\scshape i\kern-0.25em b}\kern-0.8em\TeX}}}
\patchcmd{\maketitle}{\@copyrightspace}{}{}{}

\usepackage{booktabs} 
\usepackage{url}
\usepackage{color}
\usepackage{enumitem}
\usepackage{amsmath}
\usepackage{algpseudocode}
\usepackage{multirow}
\usepackage{booktabs} 
\usepackage{caption}
\usepackage{subcaption}
\usepackage{wrapfig}
\usepackage{makecell}
\usepackage[ruled]{algorithm}

\DeclareMathOperator*{\argmin}{argmin}
\DeclareMathOperator*{\argmax}{argmax}

\usepackage{url}

\begin{document}
\title{Adversarial Item Promotion: Vulnerabilities at the Core of Top-N Recommenders that Use Images to Address Cold Start}

\author{Zhuoran Liu, Martha Larson}
\affiliation{Radboud University, Netherlands}
\email{\string{z.liu, m.larson\string}@cs.ru.nl}

\begin{abstract}
E-commerce platforms provide their customers with ranked lists of recommended items matching the customers' preferences.
Merchants on e-commerce platforms would like their items to appear as high as possible in the top-N of these ranked lists.
In this paper, we demonstrate how unscrupulous merchants can create item images that artificially promote their products, improving their rankings.
Recommender systems that use images to address the cold start problem are vulnerable to this security risk.
We describe a new type of attack, \emph{Adversarial Item Promotion} (AIP), that strikes directly at the core of Top-N recommenders: the ranking mechanism itself.
Existing work on adversarial images in recommender systems investigates the implications of conventional attacks, which target deep learning classifiers.
In contrast, our AIP attacks are embedding attacks that seek to push features representations in a way that fools the ranker (not a classifier) and directly lead to item promotion.
We introduce three AIP attacks \emph{insider attack}, \emph{expert attack}, and \emph{semantic attack}, which are defined with respect to three successively more realistic attack models.
Our experiments evaluate the danger of these attacks when mounted against three representative visually-aware recommender algorithms in a framework that uses images to address cold start.
We also evaluate potential defenses, including adversarial training and find that common, currently-existing, techniques do not eliminate the danger of AIP attacks.
In sum, we show that using images to address cold start opens recommender systems to potential threats with clear practical implications. 
\end{abstract}

\maketitle
\pagestyle{plain}

\section{Introduction}
\label{intro}

Visually-aware recommender systems~\cite{mcauley2015image, he2016vbpr, kang2017visually} incorporate image information into their ranking mechanism.
This information helps to address the challenge of cold start since it compensates for insufficient interactions associated with new users or items.
In this paper, we show how the use of image content for cold start opens visually-aware recommenders to vulnerability.

\begin{figure}[t]
\centering
\includegraphics[width=0.45\textwidth]{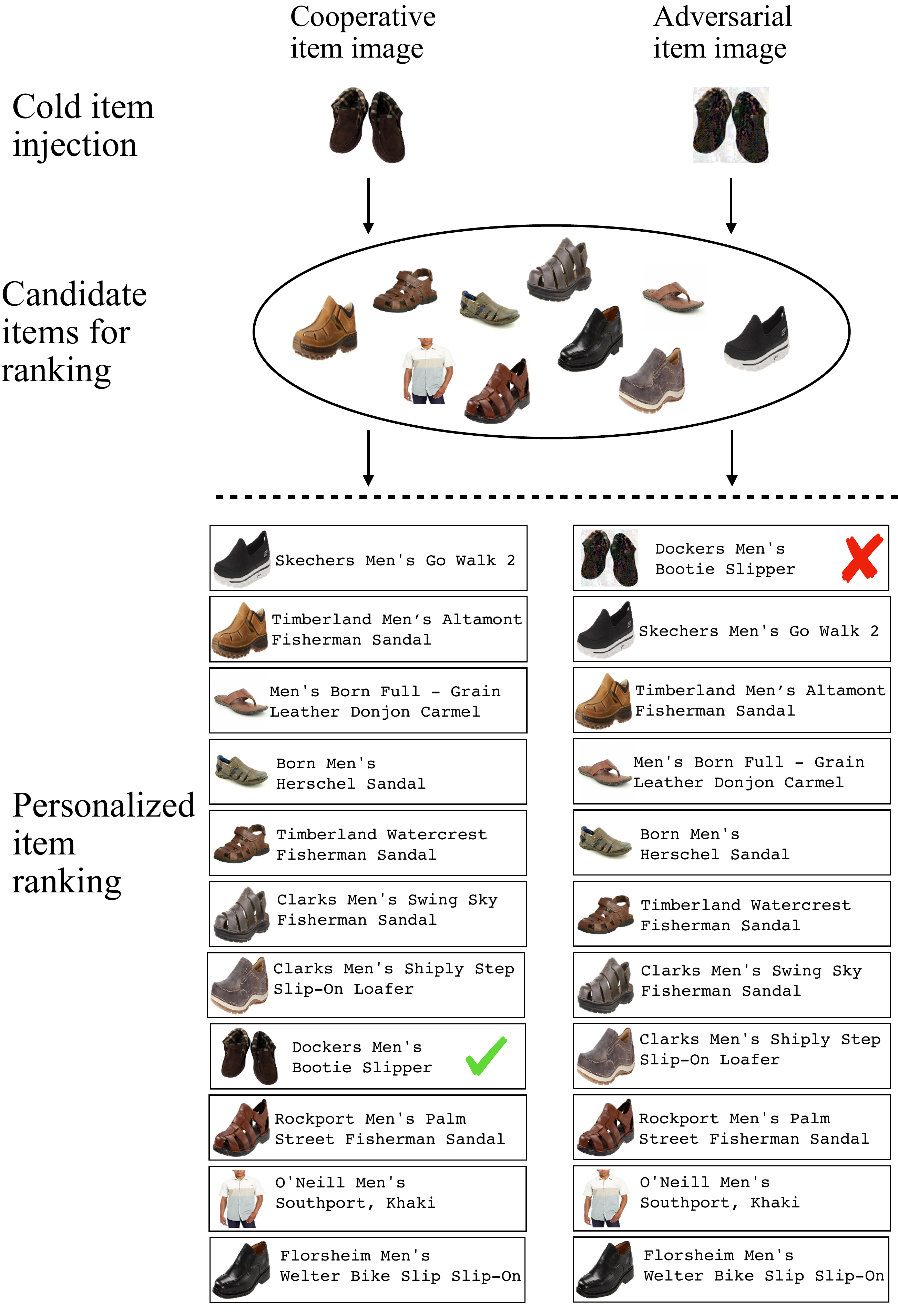}
\caption{The cooperative cold item image and its corresponding adversarial cold item image (top) are each injected into the candidate set, which is generated by a first-stage ranker (here, BPR). In the personalized ranked list generated by the visually-aware second-stage ranker (here, DVBPR) the advesarial item (right) lands much higher than the cooperative item (left). (Diagram shows a real example. The adversarial item image is generated by our INSA attack with $\epsilon = 32$ and epoch $= 5$ explained in detail in Section~\ref{sec:INSA}).}
\label{fig:diagram}
\end{figure}

The vulnerability is due to adversarial examples, which are samples deliberately designed to cause a machine learning system to make mistakes.
The computer vision community has developed an in-depth understanding of how adversarial images can be used to attack classifiers, starting with~\cite{biggio2013evasion, szegedy2013intriguing}.
Classifier-targeted adversarial images can have an impact on recommender systems that leverage image content, as has been demonstrated by TAaMR~\cite{DMM20}.
However, until now, recommender system researchers have not considered how images can be modified to create adversarial items that attack visually-aware Top-N recommender systems by directly targeting the ranker, rather than a classifier.

We expose the vulnerability of visually-aware recommender systems to adversarial items by presenting a series of attacks and by experimentally assessing the threat that they pose.
We also examine possible defenses. 
Adversarial training has been proposed in order to improve the general performance of multimedia recommender systems.
The dominant approach is currently AMR~\cite{tang2019adversarial}.
Our experiments show that AMR is not sufficient to defend against our adversarial attacks.
Further, other common defenses, such as image compression also fall short.
In sum, the vulnerabilities of visually-aware recommender systems that we investigate here are serious and require further attention of the research community.

Our work is part of the long tradition of research devoted to the security and robustness of recommender system algorithms~\cite{o2002promoting, lam2004shilling, mobasher2007toward, burke2015robust, li2016data, fang2018poisoning, christakopoulou2019adversarial, fang2020influence, tang2020revisit}. 
Most work, however, focuses on vulnerabilities related to user profiles.
Early work looked at \emph{shilling}~\cite{lam2004shilling}, which uses fake users.
Shilling was later generalized to \emph{profile injection attacks}~\cite{mobasher2007toward} or \emph{poisoning attack}~\cite{li2016data}.
In our work, in contrast, we are looking at attackers who are able to manipulate items directly, and, specifically, to choose item images.
In other words, instead of looking at profile-related attacks we are looking at an \emph{item representation attacks}.
Concretely, the risk of such attacks presents itself in the case of e-commerce platforms that sell the items of individual merchants, e.g., e-commerce or customer to customer (C2C) marketplaces.
The merchants create their own item description, including images.
We show that if such merchants act unscrupulously they can artificially promote their items and compromise the security of the recommender system.

Figure~\ref{fig:diagram} illustrates the mechanics of the attack that we consider in this paper, called an \emph{Adversarial Item Promotion} (AIP) attack. 
On the left, we see personalized item ranking for a user when the recommender system is \emph{not} under attack (i.e., the cold start item is a ``cooperative item''). 
On the right, we see the ranking when the recommender system is under attack by an unscrupulous merchant, who has used a manipulated image in an item representation (i.e., the cold start item is an “adversarial item”). 
This setup reflects the way that recommender system platform would add a certain number of cold items to the personalized ranked lists of users in order to allow the items to start accumulating interactions.  
We choose a two-stage recommender, since they are used in industry~\cite{davidson2010youtube, wang2018billion}.
With the two stage recommender, 
we ensure that the adversarial cold item is competing against selected candidate items that are already very relevant to the user.

We provide a short walk through Figure~\ref{fig:diagram}, which illustrates the attack at the level of a single cold item and a single user. 
A set of ``candidate items for ranking'' for that user has been selected using a conventional personalized Top-N recommender. 
Then the cold item is injected into that set. 
Finally, a visually-aware personalized Top-N recommender is used to rank the candidate item set before it is presented to the user. 
We see that in the case of the cooperative image (left), the cold item lands somewhere in the ranked list, but probably not at the top. 
In contrast, in the case of the adversarial image (right), the cold item lands at the top of the personalized item ranking.

The overall impact of the attack depends on the accumulated effect of the attack over all users, and not just a single instance of the attack shown in Figure~\ref{fig:diagram}.
It is important to understand that the final rank position of the cooperative vs. adversarial item will be different for each instance of the attack. 
However, in general, if the adversarial item of an unscrupulous merchant lands consistently farther towards the top of users' personalized recommendation lists than it deserves to, then, at large scale, the merchant will accrue considerable benefit.

We choose to focus on the cold-start problem because of its importance for recommender systems.
However, there is also another reason.
A straightforward, practical approach to blocking adversarial image promotion attacks on non-cold-start items is to prevent merchants from being able to change images once their items have started to accumulate interactions.
Cold start is the most important moment of opportunity for a merchant to introduce an adversarial image into a representation.
Every item starts in some way cold, and the issue particularly extreme in C2C marketplaces selling many unique items.

With this paper, we make the following contributions:
\begin{itemize}
\item We propose three Adversarial Item Promotion (AIP) attacks on the ranking mechanism of visually-aware recommender systems in a cold item scenario and experimentally assess their impact. The attacks correspond to three different levels of knowledge and we assess their impact using two real-world data sets.
\item We show that there is no easy defense against AIP attacks. The currently dominant adversarial training, as well as conventional defenses such as compression, are not sufficient to eliminate the vulnerability.
\item We release an implementation of our attacks and defenses that allows for testing and extension.\footnote{Code available at: \url{https://github.com/liuzrcc/AIP}}
\end{itemize}

This paper follows the standard procedure for security research.
First, we specify a framework including the types of attacks expected (attack models), the systems to be attacked, and a means of measuring the impact of the attacks.
Then, we propose attacks for each attack model and evaluate their success.
The systems that we attack are representative of visually-aware recommender systems, i.e., a visual feature-based similarity model (AlexRank) based on AlexNet~\cite{krizhevsky2012imagenet}, a Collaborative-Filtering (CF) model leveraging visual features (VBPR~\cite{he2016vbpr}), and state-of-the-art learning-based neural model (DVPBR~\cite{kang2017visually}).
Finally, we turn to the analysis of possible defenses and close with a conclusion and outlook.

\section{Related Work}
\subsection{Robustness of Recommender System}
In this section, we review previous work on recommender robustness.
Note that this work focused on user profiles, not image content.
O'Mahony~\emph{et al.}~\cite{o2002promoting} introduce the definition of recommender system robustness and present several attacks to identify characteristics that influence robustness.
Lam and Riedl~\cite{lam2004shilling} explore shilling attacks in recommender systems by evaluating recommendation performance under different scenarios. In particular, they find that new or obscure items are more especially susceptible to attack, and they suggest that obtaining ratings from a trusted source for these items could make them less vulnerable.
Mobasher~\emph{et al.}~\cite{mobasher2007toward} propose a formal framework to characterize different recommender system attacks, and they also propose an approach to detect attack profiles.
In ~\cite{mobasher2007toward} and~\cite{burke2015robust}, evaluation metrics, e.g., hit rate for item and prediction shift, for robustness of recommender system are discussed.
Recently, instead of model agnostic profile injection attack, many poisoning attacks that leverage exact recommendation model information are proposed.
Li~\emph{et al.}~\cite{li2016data} propose poisoning attacks to factorization-based CF algorithms by approximating the gradient based on first-order KKT conditions.
Christakopoulou and Banerjee~\cite{christakopoulou2019adversarial} propose a generative approach to generate fake user profiles to mount profile injection attack.
Fang~\emph{et al.} first propose poisoning attacks to graph-based recommender systems~\cite{fang2018poisoning}, then propose to generate fake user-item interactions based on influence function~\cite{fang2020influence}.
Tang~\emph{et al.}~\cite{tang2020revisit} propose effective transfer-based poisoning attacks against recommender system, but they discuss that their approach is less effective on cold items.
Our ``item representation attack'' is distinct from a ``profile injection attack'' or ``poisoning attack'', but both two kinds of attacks have similar impact results, namely, pushing items that have been targeted for promotion.

\subsection{Visually-aware Recommender System}
Visually-aware recommender systems incorporate visual information into their recommendation ranking mechanism.
Originally, visually-aware recommender rely on image content retrieval to make preference predictions.
Given a query image, Kalantidis~\emph{et al.}~\cite{kalantidis2013getting} gather segmentation parts and retrieve visually similar items within each of the predicted classes.
Later, semantic information of images is also incorporated to improve retrieval performance. 
Jagadeesh~\emph{et al.}~\cite{jagadeesh2014large} gather a large-scale dataset, Fashion-136K, with detailed annotations and propose several retrieval-based approaches to recommend the missing part based on query image.

Beyond image retrieval-based recommendation approaches, user-item interactions are leveraged in visually-aware recommenders.
IBR~\cite{mcauley2015image} models human notions of similarity by considering alternative or complementary items.
Later, algorithms incorporate the visual feature-based algorithms into CF model to exploit both user feedback and visual features simultaneously, e.g., VBPR~\cite{he2016vbpr} and Fashin DNA~\cite{bracher2016fashion}.
Recently, with the advances in computational resources, learning-based neural frameworks were proposed and achieve state of the art performance on fashion recommendation (DVBPR~\cite{kang2017visually}) and reciprocal recommendations (ImRec~\cite{neve2020imrec}).
In our paper, to comprehensively evaluate AIP attacks in different recommenders, we select image retrieval-based similarity model, the CF model leveraging visual features and learning-based neural model as representatives.

\subsection{Adversarial Machine Learning}
Adversarial examples are data samples that are deliberately designed in order to mislead machine learning algorithms~\cite{biggio2013evasion, szegedy2013intriguing}.
A limited amount of work, as mentioned above, has addressed adversarial images for recommender systems.
The work most closely related to our own~\cite{DMM20} looks only at classification-based issues.
Di Noia~\emph{et al.}~\cite{DMM20} propose Targeted Adversarial Attack against Multimedia Recommender Systems (TAaMR), and they use two classification-based adversarial attacks, namely Fast Gradient Sign Method (FGSM)~\cite{szegedy2013intriguing} and Projected Gradient Descent (PGD)~\cite{kurakin2016adversarial}, to evaluate two visually-aware recommender systems.
In contrast to~\cite{DMM20}, we show that the problem of adversarial examples in recommender system goes beyond the problem of classifier-targeted adversarial examples.

Adversarial training is a promising techniques to tackle adversarial examples~\cite{madry2017towards}.
As stated in Section~\ref{intro}, research on adversarial training in visually-aware recommnder systems has, until this point, focused on improving general performance.
Specifically, AMR~\cite{tang2019adversarial} aims to improve recommendation with adversarial training (cf. Section~\ref{sec:AT} for details) and considers the robustness of recommender systems perturbations in system-internal representations. 
In contrast, our goal is to investigate security vulnerability originating from an external adversary who attacks item images.
We show that simple adversarial training (i.e., AMR) is not a guarantee for robustness under AIP attacks (cf. Section~\ref{sec:AT}).

\section{Background and Framework}
\label{sec:framework}
\begin{figure}
\centering
\includegraphics[width=0.45\textwidth]{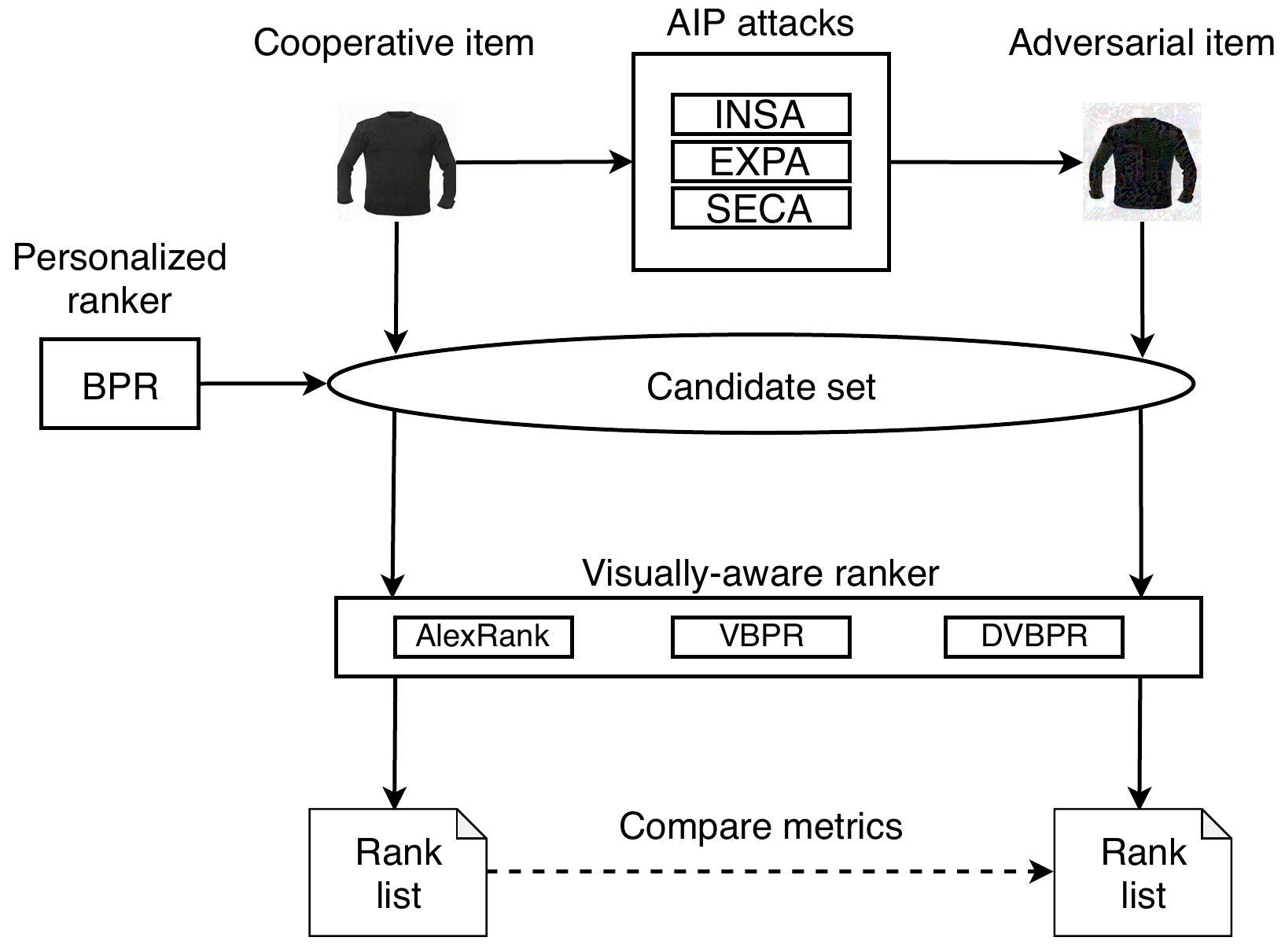}
\caption{Setup for attack and attack evaluation.}
\label{fig:diagram2}
\vspace{-0.3cm}
\end{figure}

This section introduces the background and framework in which the attack models are developed and evaluated. 
Figure~\ref{fig:diagram2} gives the overview of the setup. 
As introduced in Section~\ref{intro}, we use a two-stage approach. 
The first-stage recommender generates a personalized set of candidate items. 
For this purpose, we choose Bayesian Personalized Ranking (BPR)~\cite{rendle2012bpr}, a representative CF model that is trained on the user-item interaction data.
We use the visually-aware second-stage recommender to make a comparison between the cold start of a cooperative item and an adversarial item.

In this section, we first present our three attack models  (Section~\ref{sec:attackmodel}), which are the basis for three specific AIP attacks, INSA, EXPA, and SEMA, explained in Section~\ref{sec:attack}.
Then, we present the three representative visually-aware recommenders that we attack (Section~\ref{sec:visualrecsys}).
Finally, we explain the dimensions along which we evaluate the impact of the attacks (Section~\ref{sec:evaluation}).

\begin{table}[ht]
\centering
\caption{The attack models used to develop our AIP attacks.}
\resizebox{0.5\textwidth}{0.05\textheight}{
\begin{tabular}{l|ccc}
\toprule[1pt]
&\Huge General Knowledge&\makecell{\Huge Visual feature \\ \Huge extraction model} &\Huge Embeddings\\
\midrule[1pt]
\makecell{{\bf \Huge High-knowledge attack model}\\(\Huge  corresponds to INSA cf. Section~\ref{sec:INSA})} &&&\Huge $\times$\\ \\
\makecell{{\bf \Huge  Medium-knowledge attack model}\\(\Huge  corresponds to EXPA cf. Section~\ref{sec:EXPA})}  &\Huge  $\times$&\Huge $\times$&\\ \\
\makecell{{\bf \Huge Low-knowledge attack model}\\(\Huge  corresponds to SEMA cf. Section~\ref{sec:SEMA})}&\Huge $\times$&&\\
\bottomrule[1pt]
\end{tabular}}
\label{tbl:attackmodel}
\vspace{-0.5cm}
\end{table}

\subsection{Attack Models}
\label{sec:attackmodel}
We define three attack model following the three dimensions relevant for trustworthy recommender systems~\cite{mobasher2007toward}. 
The \emph{Intent} dimension captures the objective of the attacker.
All our models use `push' intent, i.e., attackers are merchants who want their items promoted to higher ranks in users' personalized recommendation lists.
The \emph{Knowledge} dimension captures how much information that the attacker has about the system being attacked.
Defining knowledge levels is common in adversarial machine learning research~\cite{biggio2013evasion,carlini2017adversarial}.
Our three attack models correspond to three different levels of knowledge: high, medium, and low.
The \emph{Scale} dimension captures the scope of the interference.
In all our attack models, we assume an attack with minimum scale, e.g., only a single item is attacked at any given moment.
Such a small-scale attack is not likely to be noticed.
It is clear that the harm caused by the attack will increase with scale.

Table~\ref{tbl:attackmodel} summarizes our attack models in terms of the level of knowledge involved.
For our high-knowledge attack model we assume that the attacker is an insider at the recommender system platform and has access to the user embeddings of the trained recommender model.
This scenario is not particularly realistic, but it is important because it demonstrates an upper bound for the potential damage that can be inflicted by an AIP attack.

The medium- and low-knowledge attack models are more realistic and assume that the attacker has general knowledge of the market in which the recommender system operates. 
In particular, the attacker must be able to identify (by observing sales trends or advertising) at least one item that sells well on the platform. 
We refer to this item as a \emph{hook} item.
The attack is strongest when the hook item image is an image used by the recommender, but could also be an image of the item acquired elsewhere. 
As will be explained in Section~\ref{sec:attack}, the adversarial item will use the hook item to pull itself up in the ranked list.
In the medium-knowledge attack model, in addition to general knowledge, the attacker must have access to the pre-trained visual feature extractor used by the visually-aware recommender systems. 
Recommender systems leveraging pre-trained visual feature extractor are prevalent in both academic research and industry, e.g., ~\cite{mcauley2015image, he2016vbpr, lee2017large, gomes2017boosting, prevost2018deep, sachidanandan2019designer}.
These models are often released as publicly available resources.

\subsection{Visually-aware Recommender Systems}
\label{sec:visualrecsys}
In this section, we introduce the three representative visually-aware recommender systems that we will attack in our experiments.
In particular, we first introduce the visual feature-based similarity model (AlexRank), the CF model leveraging visual features (VBPR) and the learning-based neural approach (DVBPR). We chose these recommenders because they represent the three types of commonly used visually-aware recommender systems.
\subsubsection{AlexRank}
\label{sec:alexrank}
Image content retrieval-based recommendation is a nearest neighbor approach that ranks items by visual feature similarity of product images.
Such methods are commonly used as baseline approaches in visually-aware recommender system research~\cite{kang2017visually}.
Here we use the output of the second fully-connected layer of AlexNet~\cite{krizhevsky2012imagenet} as the visual feature of item images.
Given an image of item $i$, the average Euclidean distance between the visual feature of item $i$ and all items that user $u$ has interacted with is calculated, so smaller distance means higher preference prediction. 
Equation~\ref{eq:visrank} show the calculation of similarity predictor:
\begin{equation}
\label{eq:visrank}
p_{u, i} = \sum_{j \in I_{u}^{+}} \frac{- \|\Phi_{f}(\mathbf{X}_{i}) - \Phi_{f}(\mathbf{X}_{j})\|^2}{|\mathcal{I}_{u}^{+}|},
\end{equation}
where $\mathcal{I}_{u}^{+}$ is the set of items that user $u$ has interacted with, and $\mathbf{X}_{i}, \mathbf{X}_{j}$ represent images of item i and j. 
$\Phi_{f}$ is the pre-trained model for image feature extraction.
Note that the final ranking of item $i$ for user $u$ is solely determined by preference score $p_{u, i}$.

\subsubsection{VBPR}
\label{sec:bprvbpr}
Extended from BPR~\cite{rendle2012bpr}, VBPR~\cite{he2016vbpr} incorporates visual features into the CF model.
By leveraging image features of pre-trained CNN models, VBPR improves the recommendation performance of BPR.
The preference prediction of VBPR is described in Equation~\ref{eq:VBPR}:
\begin{equation}
\label{eq:VBPR}
p_{u, i} = \alpha + \beta_{u} + \beta_{i} + \gamma_{u}^{T} \gamma_{i} + \theta_{u}^{T}(\boldsymbol{E}\Phi_{f}(\mathbf{X}_{i})),
\end{equation}
where $\theta_{u}$ is the user content embedding, and $\boldsymbol{E}$ is the parameter for visual feature. 
$\Phi_{f}(\mathbf{X}_{i})$ represents the visual feature of item $\mathbf{X}_{i}$ from pre-trained model $\Phi_{f}$, and $\alpha$, $  \beta_{u}$ and $\beta_{i}$ are user, item biases and global offset term.
$\gamma_{u}$ and $\gamma_{i}$ represent the latent embedding for user and item.
For model learning, VBPR adopts the pairwise ranking optimization framework from BPR. The training triples set $\mathcal{D}_{s}$ is described in Equation~\ref{eq:tri}:
\begin{equation}
\label{eq:tri}
\mathcal{D}_{s} = \{(u, i, j)| u \in \mathcal{U} \wedge i \in \mathcal{I}_{u}^{+} \wedge	j \in \mathcal{I} / \mathcal{I}_{u}^{+}\}
\end{equation}
where $\mathcal{U}$ and $\mathcal{I}$ represent the user and item sets, $i$ represents the interacted item, and $j$ represents the non-interacted item.
A bootstrap sampling of training triples is used for model training.
The optimization objective of VBPR is described in Equation~\ref{eq:vbpropt}:
\begin{equation}
\label{eq:vbpropt}
\underset{\Theta}{\argmin} \sum_{(u, i, j) \in \mathcal{D}_{s}} - \ln \sigma (p_{u, i} - p_{u, j}) + \lambda_{\Theta} ||\Theta||^{2}
\end{equation}
where $\Theta$ represents all model parameters and $\lambda_{\Theta}$ is the weight
of regularization term.

\subsubsection{DVBPR}
\label{sec:dvbpr}
DVBPR~\cite{kang2017visually} is a concise end-to-end model whose visual feature extractor is trained directly in a pair-wise manner.
DVBPR achieves the state-of-the-art performance on several data sets for visually-aware recommendations~\cite{kang2017visually}.
The preference prediction of DVBPR is described in Equation~\ref{eq:DVBPR}:
\begin{equation}
\label{eq:DVBPR}
p_{u, i} = \theta_{u}^{T}\Phi_{e}(\mathbf{X}_{i}), 
\end{equation}
where $\theta_{u}$ is the user content embedding, and $\Phi_{e}(\mathbf{X}_{i})$ is the item content embedding where the CNN model $\Phi_{e}$ is trained directly in a pair-wise manner.

\subsection{Attack Evaluation Dimensions}
\label{sec:evaluation}
We evaluate attacks according to the aspects of \emph{integrity} and \emph{availability} distinguished in machine learning security~\cite{barreno2010security}.
Our main concern is the ability of the attack to compromise the integrity of the recommender system, which is related to the success of the `push' intent of our attack models (cf. Section~\ref{sec:attackmodel}).
We measure the ability of our attacks to raise the rank of cold-start items such that adversarial cold start items land higher than the corresponding cooperative cold start items in users' personalized recommendation lists.
Specifically, we report the change in rank of cold-start items with prediction shift and change in hit rate (cf. Section~\ref{sec:metrics}).
In addition, we measure availability, which is related to whether the recommender system remains useful to other merchants and to customers while under attack.
For this purpose, we use `ordinary' test items. (We adopt the test items defined by the data sets.)
We calculate change in hit rate when a ordinary test item is in a candidate set with a cooperative cold item and when the same ordinary test item is in a candidate set with an adversarial cold item.

We also consider the extent to which the attack is noticeable to the human eye. 
As attacks increase in strength, perturbations become visible in images and adversarial images can be identified by experts who know what they were looking for.
However, e-commerce platforms are so large and the turnover of items so fast that it is impossible to manually vet all of the images representing items, cf. the known difficulty of filtering item collections for banned or unsafe products~\cite{berzon2019}. 
For these reason, strong attacks are quite realistic.
By focusing our experiments on strong attacks, we can evaluate the extent of the vulnerability of the system.
We also carry out additional experiments that demonstrate the effect of an attack increasing in strength from weak to strong.
In this way, we shed light on what might happen if user clicks are affected by an impression of low quality due to the presence of perturbations in images.
Our additional experiments show that the success of the attack is not dependent on a highly noticeable change to the image appearance.

\section{Adversarial Item Promotion Attacks}
\label{sec:attack}
In this section, we introduce three adversarial item promotion (AIP) attacks corresponding to the three attack models previously introduced (cf. Section~\ref{sec:attackmodel}).

\subsection{Insider Attack (INSA)}
\label{sec:INSA}
The high-knowledge attack model assumes the attacker has insider access to the user embeddings of the trained model (see Section~\ref{sec:attackmodel}).
Some visually-aware recommenders (e.g., AlexRank) only use visual embedding (feature) to build nearest neighbor-based recommender, and other recommenders (e.g., DVBPR) model the visual content embedding together with user content embedding using dot product in a CF manner.(cf. Section~\ref{sec:visualrecsys}).
For instance, in DVBPR, the inner product of user embedding $\theta_{u}$ and item embedding $\Phi_{e}(\mathbf{X}_{i})$ represents the preference of user $u$ on item $i$.

We propose an insider attack (INSA) in which the attacker can modify the embeddings of item images in order to increase the predicted preference score that solely determines the recommendation ranking.
Specifically, INSA changes item embeddings by adding perturbations $\boldsymbol{\delta}$ on the item images.
The perturbations are optimized iteratively such that the strength of preference for the item is maximized over all user profiles.
In this paper, the magnitude of $\boldsymbol{\delta}$ is restricted by $L_{\infty}$ norm, which represents the maximum value of $\boldsymbol{\delta}$ and is commonly used in computer vision research to measure imperceptibility~\cite{szegedy2013intriguing, carlini2017adversarial}.
Formally, given a product image $\mathbf{X}_{i}$ of item $i$, we optimize perturbations $\boldsymbol{\delta}$ to increase the preference $p_{u, i}$ of all users on item $i$.
The optimization objectives for different recommenders are described in Equation~\ref{eq:INSA}:

\begin{equation} 
\begin{split}
\text{\small{AlexRank}}: &\underset{\boldsymbol{\delta}}{\argmax}\sum_{u \in \mathcal{U}}\sum_{j \in I_{u}^{+}} \frac{- \|\Phi_{f}(\mathbf{X}_{i} + \boldsymbol{\delta}) - \Phi_{f}(\mathbf{X}_{j})\|^2}{|\mathcal{I}_{u}^{+}|} \\
\text{\small{VBPR}}: &\underset{\boldsymbol{\delta}}{\argmax} \sum_{u \in \mathcal{U}} \theta_{u}^{T}(\boldsymbol{E}\Phi_{f}(\mathbf{X}_{i} + \boldsymbol{\delta})) \\
\text{\small{DVBPR}}: &\underset{\boldsymbol{\delta}}{\argmax} \sum_{u \in \mathcal{U}} \exp{(\theta_{u}^{T} \Phi_{e}(\mathbf{X}_{i} + \boldsymbol{\delta}))}
\end{split}
\label{eq:INSA}
\end{equation}
$\Phi : \mathbf{X}_{i} \rightarrow \theta_{i}$ is the feature extraction or embedding model where $\theta_{i}$ represents the content embedding for item $i$.
$\theta_{u}$ represents the user content embedding.
The optimization can be implemented by mini-batch gradient descent, and it stops until certain conditions are met, e.g., it reaches certain number of iterations.

\subsection{Expert Attack (EXPA)}
\label{sec:EXPA}
The medium-knowledge attack model assumes that the attacker can select a hook (i.e., popular) item and also has access to the visual feature extraction model (see Section~\ref{sec:attackmodel}).
We propose an expert attack (EXPA) in which the attacker uses the hook item to mark the region of item space to which the adversarial item should be moved.
Specifically, the EXPA attack generates perturbations added to the cooperative item in order to create the adversarial item by decreasing the representation distance to the hook item.

Formally, generating an adversarial item image by EXPA is described in Equation~\ref{eq:EXPA}:
\begin{equation}
\label{eq:EXPA}
\underset{\boldsymbol{\delta}}{\argmin} \quad \|\Phi(\mathbf{X}_{i} + \boldsymbol{\delta}) - \Phi(\mathbf{X}_{\mathbf{hook}})\|_{2},
\end{equation}
where $\Phi$ is the feature extraction or embedding model.
The EXPA attack leverages the same mechanism as the targeted visual feature attack proposed by~\cite{sabour2015adversarial}.
The novelty of EXPA is its use of a hook image that moves the adversarial image in image space in a way that makes it rise in personalized recommendation lists.
Note that the hook image itself is not necessarily present in candidate set, which is selected by BPR, and thereby also not in the recommendation lists.
\begin{algorithm}[htb!]
\small
\caption{Adversarial Item Promotion Attack}
\label{alg:adv}
\begin{flushleft}
\algrenewcommand\algorithmicrequire{\textbf{Input:}}
\algrenewcommand\algorithmicensure{\textbf{Output:}}
\algorithmicrequire{\\$\mathbf{X}$: cold item image, $\mathbf{X}_{\mathbf{hook}}$: hook item image\\
$\boldsymbol{\delta}$: adversarial perturbations, $\epsilon$: $L_{\infty}$ norm bound\\
$\Phi$: neural network, $\theta$: user content embedding\\
$K$: number of iterations, $\mathcal{A}$: attack to mount (INSA or EXPA)\\}

\algorithmicensure{\\$\mathbf{X}^{'}$: adversarial product image}
\end{flushleft}
\begin{algorithmic}[1]
\label{algo:white-box}
\State Initialize $\boldsymbol{x}_0'\leftarrow \mathbf{X}$,\Comment\parbox[t]{.5\linewidth}{$\boldsymbol{x}_k'$ represents adversarial image in iteration $k$}\\
$\boldsymbol{\delta}\leftarrow \boldsymbol{0}$

\For {$k\leftarrow 1$ to $K$}

\If {$\mathcal{A}$ is INSA}

\State \text{AlexRank}: $\boldsymbol{\delta} \leftarrow \underset{\boldsymbol{\delta}}{\argmax}\sum_{u \in \mathcal{U}}\sum_{j \in I_{u}^{+}} \frac{- \|\Phi_{f}(\boldsymbol{x}_{k-1}' + \boldsymbol{\delta}) - \Phi_{f}(\mathbf{X}_{j})\|^2}{|\mathcal{I}_{u}^{+}|}$ \\
\State \text{VBPR}: $\boldsymbol{\delta} \leftarrow \underset{\boldsymbol{\delta}}{\argmax} \sum_{u \in \mathcal{U}} \theta_{u}^{T}(\boldsymbol{E}\Phi_{f}(\boldsymbol{x}_{k-1}' + \boldsymbol{\delta}))$ \\
\State \text{DVBPR}: $\boldsymbol{\delta} \leftarrow \underset{\boldsymbol{\delta}}{\argmax} \sum_{u \in \mathcal{U}} \exp{(\theta_{u}^{T} \Phi_{e}(\boldsymbol{x}_{k-1}' + \boldsymbol{\delta}))}$ \Comment{Eq.(\ref{eq:INSA})}
\ElsIf {$\mathcal{A}$ is EXPA}

\State $\boldsymbol{\delta} \leftarrow \underset{\boldsymbol{\delta}}{\argmin} \quad \|\Phi(\boldsymbol{x}_{k-1}' + \boldsymbol{\delta}) - \Phi(\mathbf{X}_{\mathbf{hook}})\|_{2}$ \Comment{Eq.(\ref{eq:EXPA})}

\Else

\State break

\EndIf

\State $\boldsymbol{\delta} \leftarrow
\mathrm{clip}(\boldsymbol{\delta}, -\epsilon, \epsilon)$\Comment\parbox[t]{.5\linewidth}{Make sure that the magnitude of perturbations are in pre-defined $L_{\infty}$ norm range}

\State $\boldsymbol{x}_{k}' \leftarrow \boldsymbol{x}_{k-1}' + \boldsymbol{\delta}$

\State $\boldsymbol{x}_{k}'\leftarrow \mathrm{clip}(\boldsymbol{x}_{k}'+\boldsymbol{\delta},0,1)$\Comment\parbox[t]{.5\linewidth}{Ensure perturbed image stays in valid image range}

\EndFor

\State $\boldsymbol{x}_{k}'\leftarrow\mathrm{quantize}(\boldsymbol{x}_{k}')$\Comment\parbox[t]{.5\linewidth}{Ensure $\boldsymbol{x}_{k}'$ is valid in the 8-bit image format}

\State \Return $\mathbf{X^{'}} \leftarrow\boldsymbol{x}_{k}'$ is the adversarial item image 
\end{algorithmic}
\end{algorithm}

Algorithm~\ref{alg:adv} describes the process to generate adversarial product images with INSA and EXPA. 
$\mathbf{X}_{i}$ is the original image of cold item, and $\mathbf{X}_{\mathbf{hook}}$ is the hook item.
$\Phi$ is the neural network that extracts embeddings or features from the image content.
Our aim is to find perturbations $\boldsymbol{\delta}$ that could increase the personalized preference predictions by optimization through all user content embeddings (INSA) or targeting a hook item (EXPA). 
The magnitude of perturbations can be adjusted by of $\boldsymbol{\delta}$.
To make sure that output images are valid in standard image encoding format, a clip function restricts adversarial item image in range [0, 1], and a quantization function ensures that the output image can be saved in the 8-bit format.
The resulting adversarial image $\mathbf{X}^{'}$ is the summation of the original image and the clipped perturbations.

\subsection{Semantic Attack (SEMA)}
\label{sec:SEMA}
The low-knowledge attack model assume nothing beyond general knowledge needed to choose hook items (see Section~\ref{sec:attackmodel}).
We propose a semantic attack (SEMA) uses the semantic content of the image, i.e., what is shown in the image, in order to achieve the promotion of items.
The attack differs considerably from INSA and EXPA, which add perturbations to existing images without changing what the images depict.

\begin{figure*}
\centering
\includegraphics[width=\textwidth]{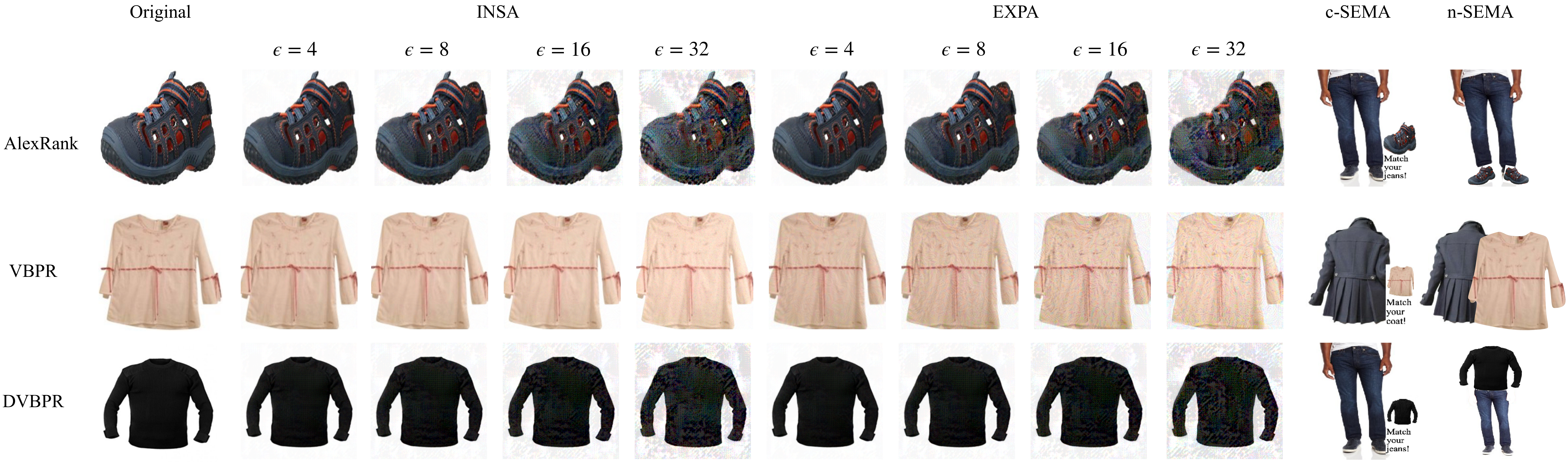}
\caption{Examples of adversarial item images by different approaches. INSA attack is based on Equation~\ref{eq:INSA}. EXPA, c-SEMA and n-SEMA select a popular item, e.g., Levi's 501 jeans, as the hook item. c-SEMA applies simple co-depiction approach, and n-SEMA incorporates the target item in a more natural way. More details about the influence of $\epsilon$ on recommendation performance can be found in Section~\ref{sec:influence_hype}}
\label{fig:imageexamples}
\end{figure*}

Figure~\ref{fig:imageexamples} c-SEMA illustrates the semantic attack that we will test here, which we call composition semantic attack (c-SEMA).
With c-SEMA, the attacker creates an adversarial image by editing the original image into the hook item image as an inset.
Here, the c-SEMA attack is promoting a pair of shoes and the hook item is the jeans.
A text (here, “match your jeans/coat”) can be included to contribute to the impression that the composite image is a standard attempt to raise the interest of potential customers.

Figure~\ref{fig:imageexamples} n-SEMA shows another type of semantic attack, which we call natural semantic attack (n-SEMA).
Here, the integration of the hook item is natural.
n-SEMA images can be created in a photo studio or a professional photo editor.
We do not test them here, since creation is time consuming and we are using a cold test set of 1000 item images.
However, an unscrupulous merchant would have the incentive to invest the time to create n-SEMA images.
One highly successful adversarial item image could already lead to increased buyers and increased profit.

The semantic attack is particularly interesting for two reasons. 
SEMA achieves the change in image embeddings needed to push an adversarial image close to a hook image in image space by manipulating the depicted content of the image.
First, this means that there are no limits on the quality of a SEMA adversarial image.
Contrast c-SEMA and n-SEMA with the INSA and EXPA photos in Figure~\ref{fig:imageexamples}.
The item is visible in the image, and consumers who decide to purchase the product will not find that they have been misled.
However, not all of the images are crisp, and stronger attacks introduce artifacts affecting the perceived image quality. 
Second, the impact of a SEMA image attack is not dependent on the algorithm used by the recommender systems.
In fact, SEMA images can effectively attack any recommender system using visual features, and not just systems using neural embedding as studied here.

\section{Experimental Setup}
\label{sec:exp_setup}
In this section, we first introduce data sets used for our experiments (Section~\ref{sec:data}) and introduce the details of evaluation setup including metrics (Section~\ref{sec:metrics}).
Then, we describe implementation of experiments (Section~\ref{sec:imp_detais}).

\subsection{Data}
\label{sec:data}
\subsubsection{Data statistics}

\begin{table}[]
\centering
\newcommand{\tabincell}[2]{\begin{tabular}{@{}#1@{}}#2\end{tabular}}
\caption{Statistics of the data sets}
\vspace{-0.2cm}
\begin{tabular}{lccc}
\toprule
& \tabincell{c}{\#Users}               & \tabincell{c}{\#Items}             & \tabincell{c}{\#Interactions}\\
\midrule
Amazon Men       &34244      &110636 &254870         \\
Tradesy.com  &33864       &326393 &655409  \\
\bottomrule
\end{tabular}
\label{tbl:datasta}
\vspace{-0.5cm}
\end{table}
We perform our experiments on two data sets: Men's Clothing in Amazon.com and Tradesy.com, which are publicly available and widely used in visually-aware recommender system research.
The statistics of the two data sets are described in Table~\ref{tbl:datasta}. Men's Clothing category is an important subset of Amazon.com data set, where the effectiveness of visual features have been validated in previous work~\cite{mcauley2015image, he2016vbpr, kang2017visually}.
Tradesy.com is a C2C second-hand clothing trading website where users can buy and sell fashion items.
The nature of the Tradesy.com inventory makes visually-aware cold item recommendation crucial, because its `one-off' characteristics.
For both datasets, one descriptive image is available for each item, and we follow the protocol of~\cite{kang2017visually} and treat users' review histories as implicit feedback.
For each user, one item is selected among all interacted items as the test item, so we have the same number of test items as the number of users.

\subsubsection{Cold test item election}
\label{sec:eval_set}
To validate the effectiveness of the attacks in cold start scenario, in each of Amazon Men and Tradesy.com data sets, we randomly select 1000 cold test items that no user has interacted with and leave them out as the cold test set.
These cold items are excluded from the training process.
Later, they are injected as cold-start items to the candidate item set before feeding into the visually-aware ranker.

\subsection{Evaluation Metrics}
\label{sec:metrics}
The change in rank of cold-start items is measured by the prediction shift and the change in hit rate ($\mathit{HR}@N$), following the evaluation metrics for top-N recommender system robustness~\cite{mobasher2007toward} and~\cite{burke2015robust}.
Equation~\ref{eq:coldtestshift} defines the average prediction shift $\Delta_{p_{i}}$ for item i and also the mean average prediction shift for a set of test items $\Delta_{\mathbf{set}}$. $p'_{u, i}$ is the post-attack predictor score and $p_{u, i}$ is the original predictor score for item $i$.
$\mathcal{I}_{\mathbf{test}}$ represents the set of test items.
\begin{equation}
\begin{aligned}
&\Delta_{p_{i}}  = \sum_{u \in \mathcal{U}} \frac{(p'_{u, i} - p_{u, i})}{|\mathcal{U}|}
&\Delta_{\mathbf{set}} = \sum_{i \in \mathcal{I}_{\mathbf{test}}}\frac{\Delta_{p_{i}}}{|\mathcal{I}_{\mathbf{test}}|},
\end{aligned}
\label{eq:coldtestshift}
\end{equation}
Equation~\ref{eq:hr} defines the average hit rate ${HR}_i @N$ for item $i$ in terms of $H_{u, i}@N$ for item $i$ for user $u$. The mean average hit rate $\textit{HR}@N$ for test items averages ${HR}_i @N$ over the test set. $\Delta_{\textit{HR}@N}$ is the change in hit rate, where $\textit{HR}^{'}@N$ is the post-attack hit rate. 
\begin{equation}
 \begin{aligned}
&\textit{HR}_i @N = \sum_{u \in \mathcal{U}} \frac{H_{u, i}@N}{|\mathcal{U}|}
&\textit{HR} @N = \sum_{i \in \mathcal{I}_{\mathbf{test}}}\frac{\textit{HR}_i @N}{|\mathcal{I}_{\mathbf{test}}|} \\
&\Delta_{\textit{HR}@N}  = \sum_{i \in \mathcal{I}_{\mathbf{test}}}\frac{\textit{HR}_i^{'} @N - \textit{HR}_i@N}{|\mathcal{I}_{\mathbf{test}}|}
\end{aligned}
\label{eq:hr}
\end{equation}
It is important to note that low metric values can still result in large impact due to the large number of users involved.
For example, in Amazon Men, an increase of $0.01$ on $\textit{HR}@5$ means that adversarial cold items are pushed into the top-5 list of about 340 users.

\subsection{Implementation}
\label{sec:imp_detais}
In the first stage, we use BPR to generate a candidate set of top 1000 items that are selected by personalized preference ranking.
Note that we use a set, which means that the original rank order is not taken into account by the visually-aware ranker.
Then we inject the ordinary test item in the top 1000 candidate set and get a set of 1001 items for each user.
To compare before and after the attack, we inject one cooperative cold item or its corresponding adversarial cold item in the candidate set.
So, for each cold-start item, we have two sets of 1002 items, which includes one test item and one cooperative (or adversarial) cold item.
We use our three visual ranking models, AlexRank, VBPR, and DVBPR (see Section~\ref{sec:visualrecsys}), to rank the 1002 items and evaluate with respect to both integrity and availability (see Section~\ref{sec:evaluation}).
\subsubsection{Model training}
We implement BPR, AlexRank, VBPR, and DVBPR in PyTorch~\cite{paszke2017automatic}.
For the first stage model BPR, we set the number of factors to 64.
Stochastic Gradient Desecent (SGD) is used to optimize BPR with learning rate 0.01 for Amazon Men and 0.5 for Tradesy.com, where the weight decay for L2 penalty is set to 0.001 on both data sets.
The feature dimension of AlexRank is 2048, and the embedding length of both VBPR and DVBPR is 100.
A grid search of learning rate in $\{0.1, 0.01, 0.001, 0.005\}$ and weight decay in $\{0.001, 0.0001, 0.0001\}$ is conducted for both VBPR and DVBPR to select hyperparameters, and we select the model with best validation performance.

\subsubsection{AIP attacks}
If not specifically mentioned, the maximal size of perturbations $\epsilon$ for AIP attacks is set to $32$.
In INSA, the number of epochs is set as $10$ to control the attacking time, and our implementation takes about 2 hours to generate 1000 adversarial item images on a single NVIDIA RTX 2080Ti GPU.
We use the Adam optimizer with the learning rate of 0.001 for DVBPR and 0.0001 for both VBPR and AlexRank.
In EXPA, the hook items are ``Levi's Men's 501 Original Fit Jean'' in Amazon Men and a gray coat in Tradesy.com.
These two products are most commonly interacted items in training data of these two data sets.
Recall, however, that hooks can be chosen without direct access to interaction statistics.
We use a Adam optimizer with a learning rate of 0.01 in EXPA, and the number of iterations is set as 5000.
In c-SEMA, we resize the hook item image and paste it on the right side of the cooperative item image as shown in Figure~\ref{fig:imageexamples}. 
To make the combination more natural, we also add a text description. 
More implementation details can be found in our released code.

\section{Experimental Results}
\label{sec:exp_results}
In this section, we carry out experimental analysis of our attacks using two real-world data sets (Section~\ref{sec:attack_eval}) and also investigate the influence of hyperparameter selections (Section~\ref{sec:influence_hype}).
Finally, we analyze and discuss classification-based attack (Section~\ref{sec:taamr}).

\subsection{Attack Evaluation}
\label{sec:attack_eval}
\begin{table*}[!htb]
\caption{Absolute mean average prediction shifts of adversarial cold items $|\Delta_{\text{set}}|$, where $\uparrow$ represents positive prediction shift (rank increased) and $\downarrow$ represents negative prediction shift (rank decreased). Positive shift means a successful attack.}
\vspace{-0.1cm}
\newcommand{\tabincell}[2]{\begin{tabular}{@{}#1@{}}#2\end{tabular}}
\renewcommand{\arraystretch}{1}
\begin{center}
\resizebox{0.65\textwidth}{!}{
\begin{tabular}{l|ccc|ccc|ccc}
\toprule[1pt]
&\multicolumn{3}{c|}{AlexRank}&\multicolumn{3}{c|}{VBPR}&\multicolumn{3}{c}{DVBPR}\\
\cline{2-10}
&INSA&EXPA&c-SEMA&INSA&EXPA&c-SEMA&INSA&EXPA&c-SEMA\\
\hline
Amazon Men &$\uparrow$16.13&$\uparrow$15.94&$\uparrow$11.1&$\uparrow$3.27&$\uparrow$3.16&$\uparrow$0.88&$\uparrow$13.54&$\uparrow$4.80&$\uparrow$4.82\\
\hline
Tradesy.com &$\uparrow$26.89&$\downarrow$0.67&$\downarrow$3.44&$\downarrow$0.79&$\uparrow$1.60&$\uparrow$1.45&$\uparrow$3.64&$\uparrow$1.89&$\uparrow$1.19\\

\bottomrule[1pt]
\end{tabular}
    }
        
\end{center}

\label{tab:ps}
\end{table*}

We mount AIP attacks and assess their effects with respect to our metrics (cf.  Section~\ref{sec:eval_set}).
Table~\ref{tab:ps} shows the absolute mean average prediction shift $|\Delta_{\text{set}}|$ for adversarial cold items vs. cooperative items.
The upwards arrow represents a positive prediction shift, meaning that the attack has successfully promoted the item.
We see that for nearly all combinations of AIP attack and visually-aware recommender system the attack is successful.
The high-knowledge attack, INSA, achieves a larger shift than EXPA and c-SEMA.
c-SEMA is surprisingly successful given the very minimal amount of knowledge that it requires.
Note that it is only meaningful to compare the size of the prediction shift for the same recommender system and the same data set.
A negative mean average prediction shift does not necessarily mean that an attack is unsuccessful since it is the rank position and not the preference prediction score that translates into benefit for the attacker.
For this reason, we go on to examine hit rate related metrics.

\begin{table*}[!htb]
\caption{Average $\mathit{HR}@5$ of cooperative (adversarial) cold item and ordinary test item in AlexRank, VBPR, DVBPR and adversarial training-based AMR. Three attacks, INSA, EXPA, and c-SEMA, are evaluated on Amazon Men and Tradesy.com data sets. $^{**}$ indicates that the increase/decrease over ordinary test/cooperative cold test are statistically significant ($p < 0.01$).}
\vspace{-0.1cm}
\parbox[b]{.742\linewidth}{
\subcaption{AlexRank, VBPR, and DVBPR.}
\newcommand{\tabincell}[2]{\begin{tabular}{@{}#1@{}}#2\end{tabular}}
\renewcommand{\arraystretch}{1}
\begin{center}
\resizebox{!}{0.08\textheight}{
\begin{tabular}{l|l|l !{\vrule width 1pt}c|c|c|c|c|c}
\toprule[1pt]
&&&\multicolumn{3}{c|}{Amazon Men}&\multicolumn{3}{c}{Tradesy.com}\\
\midrule[1pt]

\textbf{Attack method}&\textbf{Dimension}&\textbf{Test set}
&AlexRank
&VBPR
&DVBPR
&AlexRank
&VBPR
&DVBPR\\

\hline
\multirow{2}{*}{$-$}&$-$
&Cooperative cold test
&$0.0011$
&$0.0017$
&$0.0010$
&$0.0003$
&$0.0022$
&$0.0011$
\\
&$-$&Ordinary test 
&$0.0187$
&$0.0191$
&$0.0195$
&$0.0338$
&$0.0241$
&$0.0234$
\\

\hline
\multirow{2}{*}{\textbf{INSA}}&Integrity
&Adversarial cold test 
&$0.8296^{**}$
&$0.2678^{**}$
&$0.7211^{**}$
&$0.9399^{**}$
&$0.0783^{**}$
&$0.2711^{**}$
\\
&Availability&Ordinary test  
&$0.0140^{**}$
&$0.0190^{**}$
&$0.0170^{**}$
&$0.0284^{**}$
&$0.0224^{**}$
&$0.0228^{**}$
\\

\hline
\multirow{2}{*}{\textbf{EXPA}}&Integrity
&Adversarial cold test
&$0.0036^{**}$
&$0.0150^{**}$
&$0.0646^{**}$
&$0.0002^{**}$
&$0.0048^{**}$
&$0.0384^{**}$
\\
&Availability&Ordinary test 
&$0.0188^{**}$
&$0.0198^{**}$
&$0.0192^{**}$
&$0.0338$
&$0.0228^{**}$
&$0.0232^{**}$
\\

\hline

\multirow{2}{*}{\textbf{c-SEMA}}&Integrity
&Adversarial cold test
&$0.0017^{**}$
&$0.0027^{**}$
&$0.0181^{**}$
&$0.0000^{**}$
&$0.0048^{**}$
&$0.0005^{**}$
\\

&Availability
&Ordinary test 
&$0.0187^{**}$
&$0.0198^{**}$
&$0.0194^{**}$
&$0.0338$
&$0.0227^{**}$
&$0.0234$\\
\bottomrule[1pt]
\end{tabular}
    }
        
\end{center}}
\parbox[b]{.2\linewidth}{
\subcaption{AMR.}
\renewcommand{\arraystretch}{1}
\begin{center}
\resizebox{!}{0.08\textheight}{
\begin{tabular}{c|c}
\toprule[1pt]
Amazon Men&Tradesy.com\\
\midrule[1pt]
\multicolumn{2}{c}{AMR}\\

\hline
$0.0013$&$0.0013$\\
$0.0289$&$0.0381$\\
\hline
$0.7476^{**}$&$0.2858^{**}$\\
$0.0238^{**}$&$0.0362^{**}$\\
\hline
$0.0185^{**}$&$0.0039^{**}$\\
$0.0289^{**}$&$0.0381^{**}$\\

\hline
$0.0007^{**}$&$0.0047^{**}$\\
$0.0289^{**}$&$0.0381^{**}$\\

\bottomrule[1pt]
\end{tabular}}
\end{center}
}

\label{tab:huge}
\end{table*}

\begin{table*}[!htb]
\caption{Change in average $\mathit{HR}@5$ before and after AIP attacks: Cold items (attack is successful if $\mathit{HR}@5$ rises).}
\vspace{-0.1cm}
\newcommand{\tabincell}[2]{\begin{tabular}{@{}#1@{}}#2\end{tabular}}
\renewcommand{\arraystretch}{1}
\parbox[b]{.75\textwidth}{
\subcaption{AlexRank, VBPR, and DVBPR.}
\begin{center}
\resizebox{!}{0.047\textheight}{
\begin{tabular}{l!{\vrule width 1pt}ccc|ccc}
\toprule[1pt]
&\multicolumn{3}{c|}{Amazon Men}&\multicolumn{3}{c}{Tradesy.com}\\
\cline{2-7}
&AlexRank&VBPR&DVBPR&AlexRank&VBPR&DVBPR\\
\midrule[1pt]
INSA adversarial cold vs. cooperative cold
&$\uparrow 0.8285$&$\uparrow 0.2661$&$\uparrow 0.7201$
&$\uparrow 0.9396$&$\uparrow 0.0761$&$\uparrow 0.2700$\\
EXPA adversarial cold vs. cooperative cold
&$\uparrow  0.0025$&$\uparrow 0.0133$&$\uparrow 0.0636$
&$\downarrow 0.0001$&$\uparrow 0.0026$&$\uparrow 0.0373$\\
c-SEMA adversarial cold vs. cooperative cold
&$\uparrow 0.0006$&$\uparrow 0.0010$&$\uparrow 0.0171$
&$\downarrow 0.0003$&$\uparrow 0.0026$&$\downarrow 0.0006$\\

\bottomrule[1pt]
\end{tabular}
    }
\end{center}}
\parbox[b]{.2\textwidth}{
\subcaption{AMR.}
\begin{center}
\resizebox{!}{0.047\textheight}{
\begin{tabular}{c|c}
\toprule[1pt]
Amazon Men&Tradesy.com\\
\cline{1-2}
AMR&AMR\\
\midrule[1pt]
$\uparrow 0.7463$
&$\uparrow 0.2845$\\

$\uparrow 0.0172$
&$\uparrow 0.0026$\\

$\downarrow 0.0006$
&$\uparrow 0.0034$\\

\bottomrule[1pt]
\end{tabular}
    }
\end{center}

}
\label{tab:hrcompare}
\vspace{-0.3cm}
\end{table*}

Table~\ref{tab:huge}(a) presents results in terms of the mean average hit rate $\mathit{HR}@5$.
The first two rows report the original situation: the hit rate for the cooperative (i.e., not adversarial) version of the cold items and the hit rate for the ordinary test items in the case that no adversarial items have been added to the candidate set.
The rest of the table reports on attacks.
Cases marked with $^{**}$ indicate a statistically significant difference between the original situation and the case of the attack (item-level paired sample t-test $p < 0.01$).

First, we consider ``Integrity", namely, the success of the attacks in pushing items.
It can be observed in Table~\ref{tab:huge} that in nearly all cases the hit rate for the adversarial version of the cold items exceeds that of the cooperative version of the cold items, meaning that all AIP attacks are generally effective.
For INSA, the impact of the attack is dramatic. 
The cooperative cold item makes it to one of the top-5 position in the lists of only 35 users (averaged over the three recommender system), but after the attack, the adversarial cold item makes it into the top-5 position of over 20,000 users.
Since INSA uses the most knowledge, it is not surprising that it is the most effective attack.
However, even with much less knowledge, both EXPA and c-SEMA pose serious threats.
For example, for Amazon Men, 
c-SEMA pushes cold items into the top-5 list of 582 users in the case of DVBPR.
We also calculated $\mathit{HR}@10$ and $\mathit{HR}@20$ for all conditions.
These are not reported here since the trends were overwhelming the same as for $\mathit{HR}@5$. 

Next, we turn to discuss ``Availability", namely, 
the extent to which promotion occurs at the
expense of other items, which at scale can impede the functioning of the entire recommender system.
In Table~\ref{tab:huge}, we see that INSA has a strong impact on availability than EXPA and c-SEMA.
Note, however, that the different performance of EXPA and c-SEMA on Amazon Men and Tradesy.com is not solely attributable to the adversarial attack itself, since the selection of the hook item also has an impact.
For Tradesy.com, the $\mathit{HR}@5$ for the selected hook item (which is a gray coat) is originally rather low, so after EXPA or c-SEMA, the rank cannot increase dramatically.
In general, we observe that AIP attacks are more damaging to integrity than to availability.
However, in real-world situations it would be important to study cases involving the simultaneous presence of multiple adversarial items.

\begin{figure*}
\centering
\includegraphics[width=0.8\textwidth]{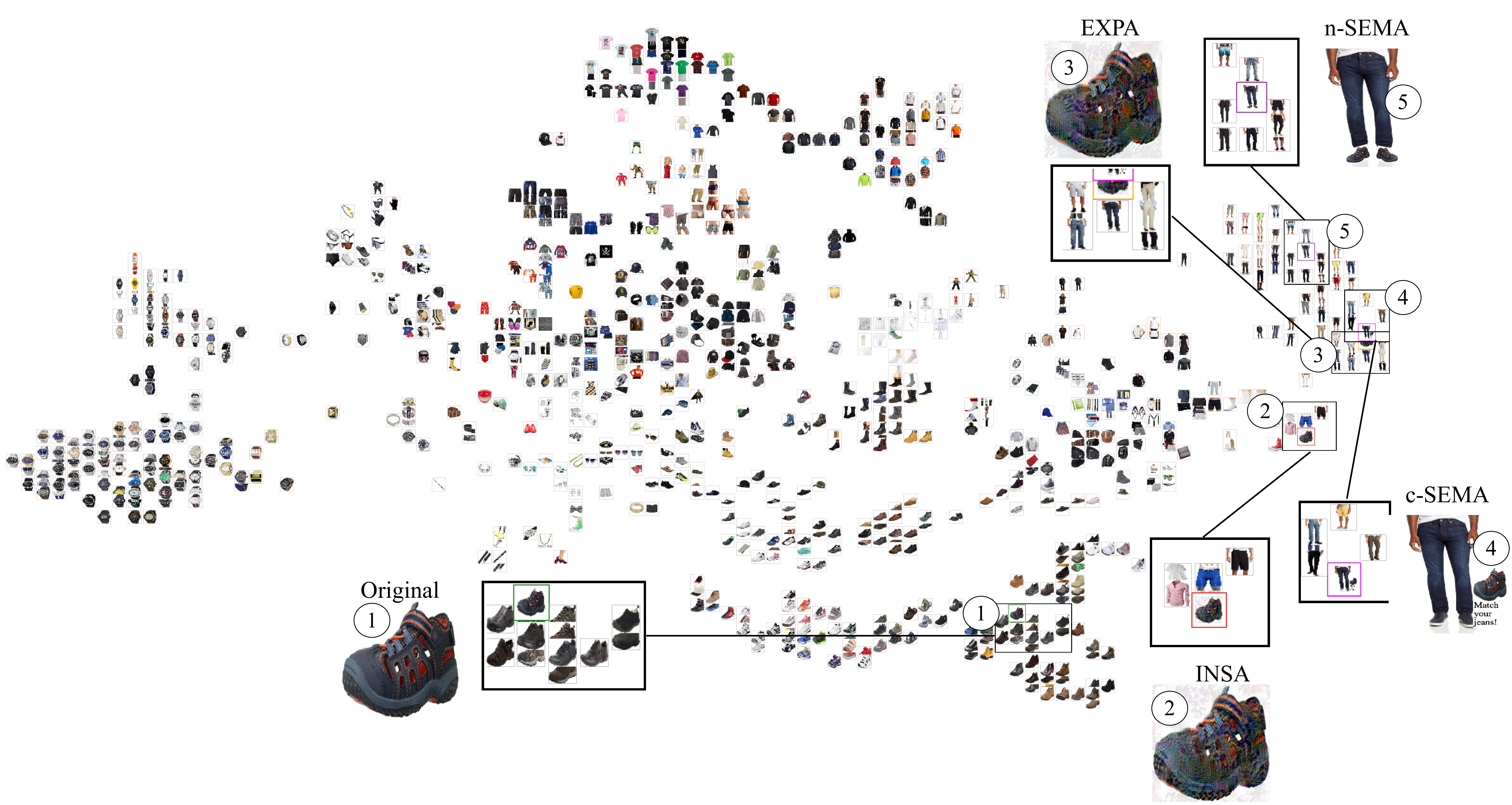}
\caption{2-D visualization of item embedding by pre-trained DVBPR model (\#factors = 100) for randomly selected 1000 items in Amazon Men data set, cooperative item ({\large \textcircled{\small 1}}) embedding and its corresponding adversarial item embedding. Attacks include generated images by INSA ({\large \textcircled{\small 2}}), EXPA ({\large \textcircled{\small 3}}), c-
SEMA({\large \textcircled{\small 4}}) and n-
SEMA({\large \textcircled{\small 5}}).}
\label{fig:tsne}
\end{figure*}

In order to directly illustrate the magnitude of the impact of AIP attacks on integrity, we report the change in mean average hit rate in Table~\ref{tab:hrcompare}(a).
Again, the expected large effect of INSA can be observed.
Also, again, the c-SEMA attack is surprisingly effective, given the minimal knowledge it involves.
Remember that the recommender system platforms we are concerned about are enormous, and even a small boost in average rank of the magnitude of that afforded by c-SEMA could translate in to a substantial increase in interactions and profit.

Figure~\ref{fig:tsne} shows a 2-D visualization (using t-SNE~\cite{maaten2008visualizing}) of the image space defined by the DVBPR item embeddings. 
It allows us to directly observe the influences of different AIP attacks.
The position of the original cold image (i.e., a cooperative image) is shown by $\large \textcircled{\small 1}$.
We can see that it is positioned next to items with which it is visually similar.
The attacks move this image to the other positions. 
The hook item is a pair of jeans.
We can see that EXPA, c-SEMA, and n-SEMA push the cold item to a cluster related to the hook item.
Note that it is difficult to reason about the position of INSA, since it is optimized with respect to all user embeddings.
Here the n-SEMA image is manually generated using photo-editing software, in particular, after cropping and rotation, the shoes to be promoted are pasted on the hook image.
However, recall that in the real world an image could easily be taken of a model wearing both items.

\subsection{Influence of Hyperparameters}
\label{sec:influence_hype}
Choice of the hyperparameters can influence the impact of the attack.
Here, we take a look at the two most important hyperparameters, embedding length and attack strength ($\epsilon$).

\begin{figure}[htb]
\centering
\begin{subfigure}{.22\textwidth}
  \centering
  \includegraphics[width=0.9\textwidth]{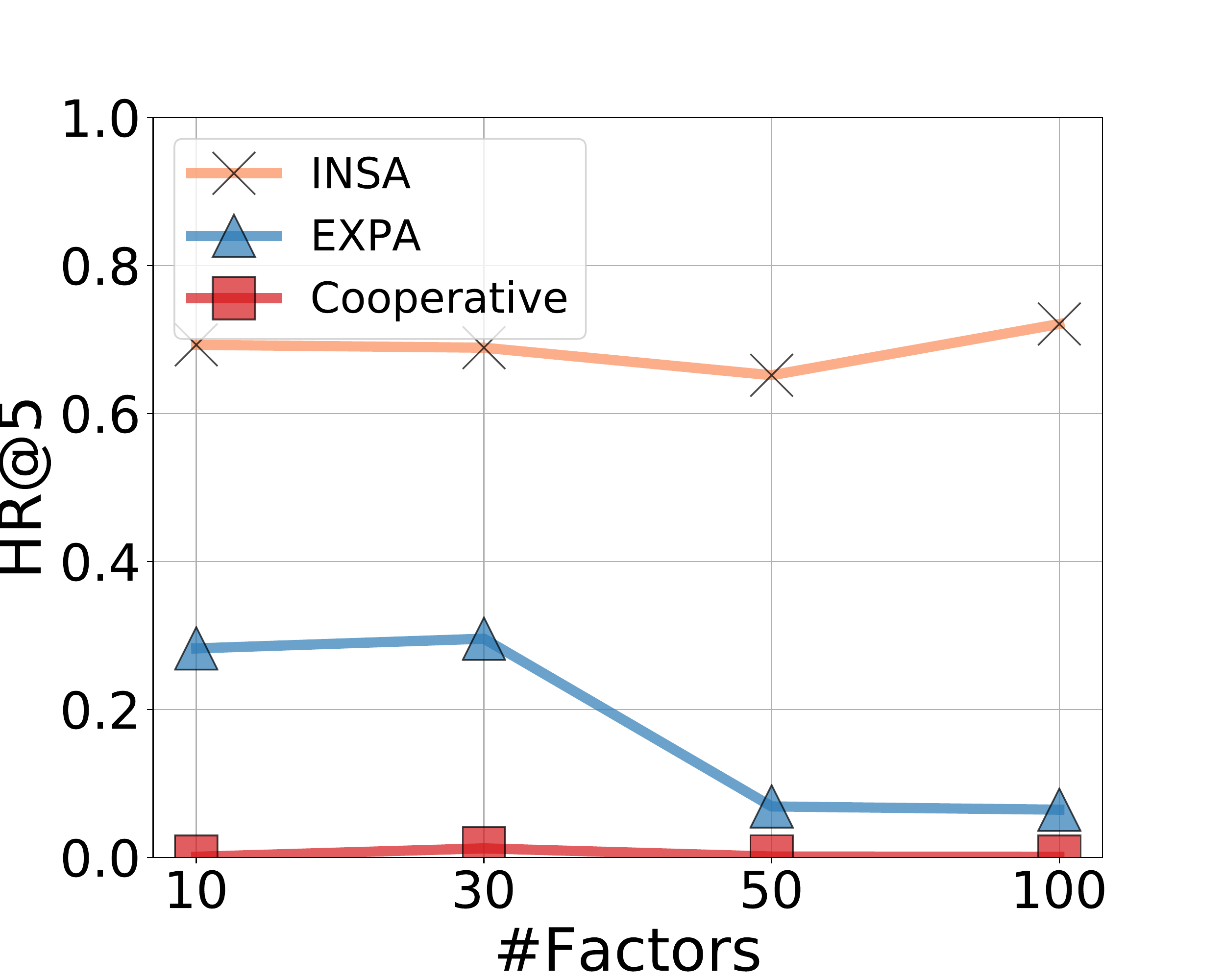}
  \caption{DVBPR (Amazon Men)}
  \label{fig:sub1}
\end{subfigure}\hfil%
\begin{subfigure}{.22\textwidth}
  \centering
  \includegraphics[width=0.9\textwidth]{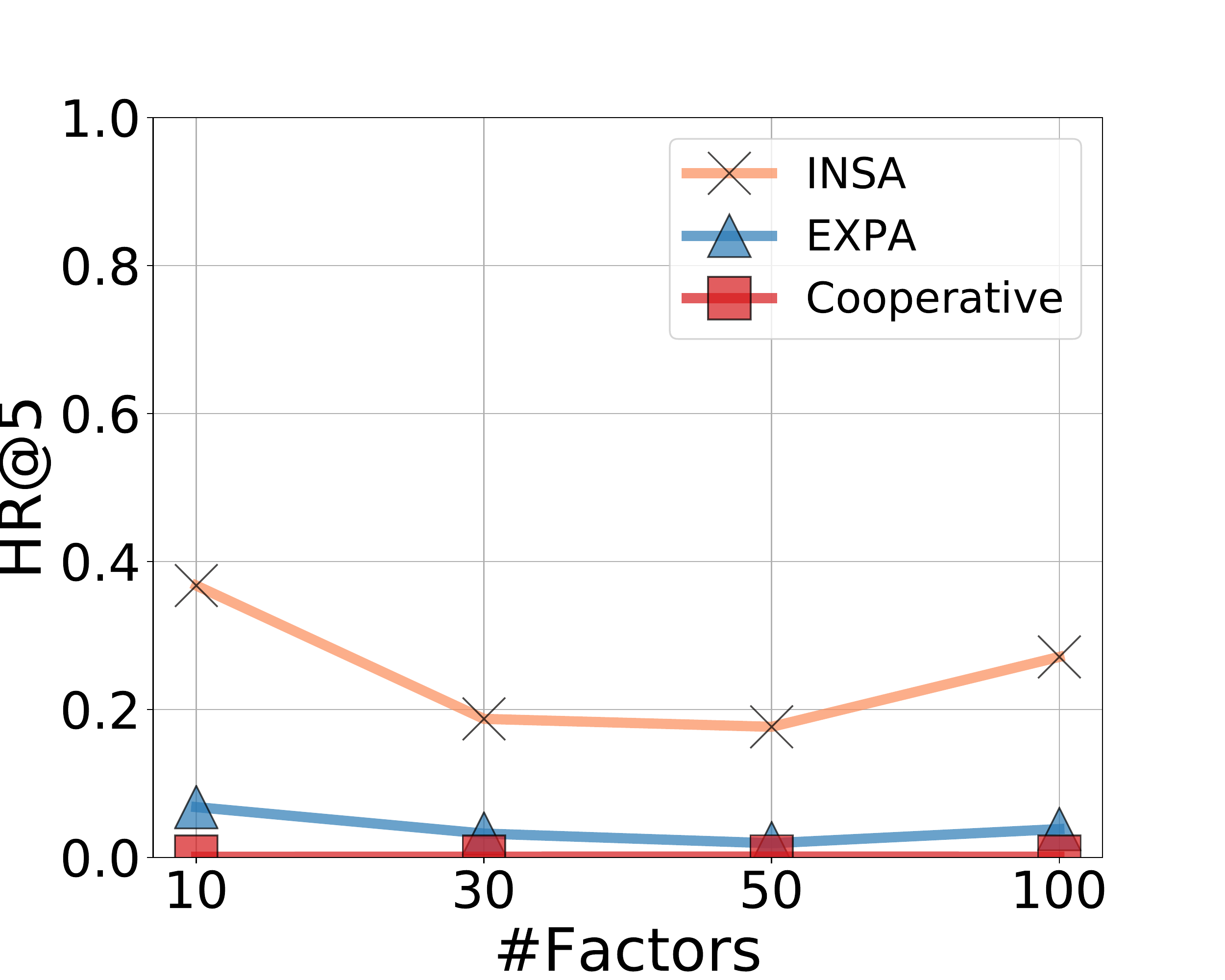}
  \caption{ DVBPR (Tradesy.com)}
\end{subfigure}\hfil
\begin{subfigure}{.22\textwidth}
  \centering
  \includegraphics[width=0.9\textwidth]{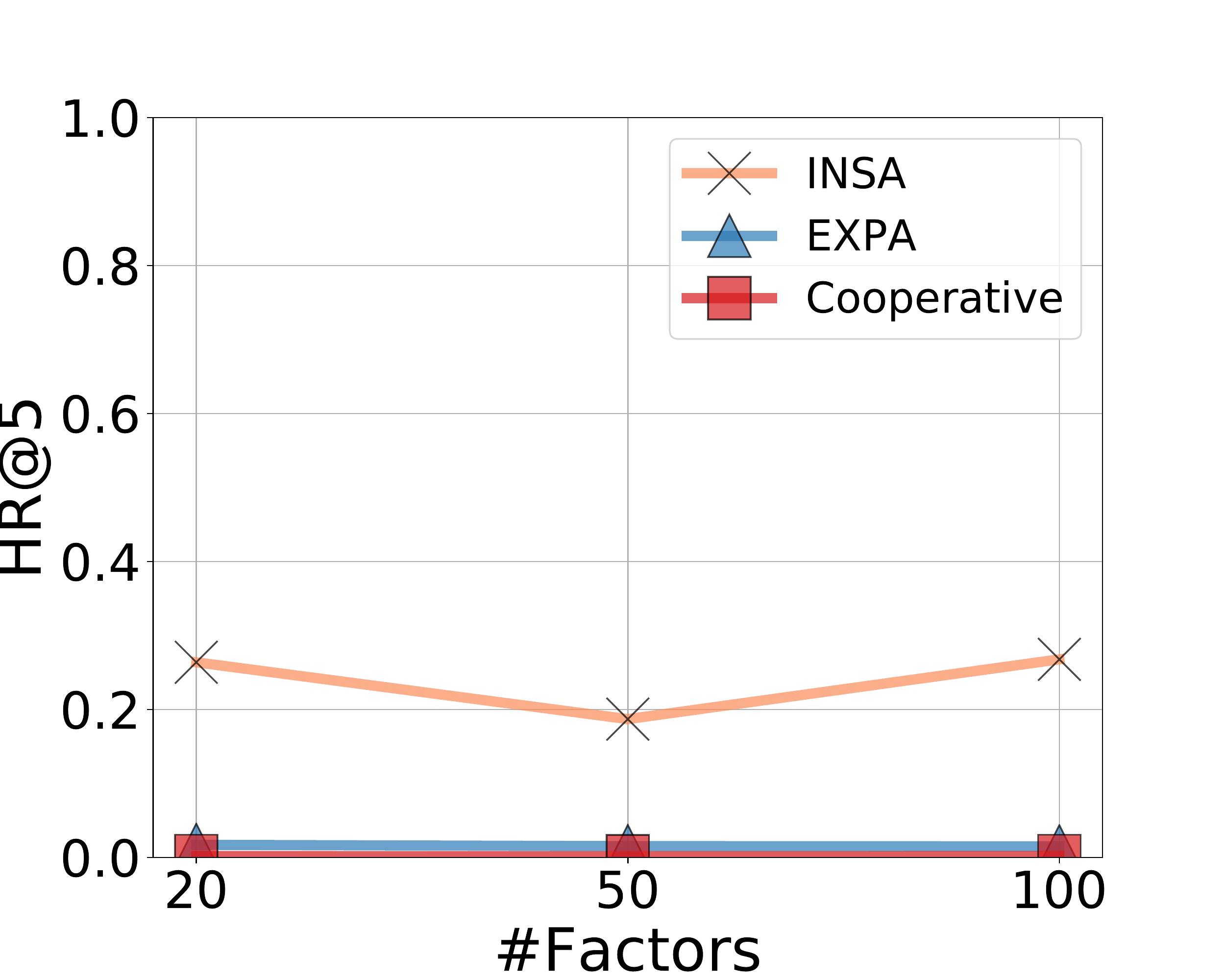}
  \caption{VBPR (Amazon Men)}
\end{subfigure}\hfil
\begin{subfigure}{.22\textwidth}
  \centering
  \includegraphics[width=0.9\textwidth]{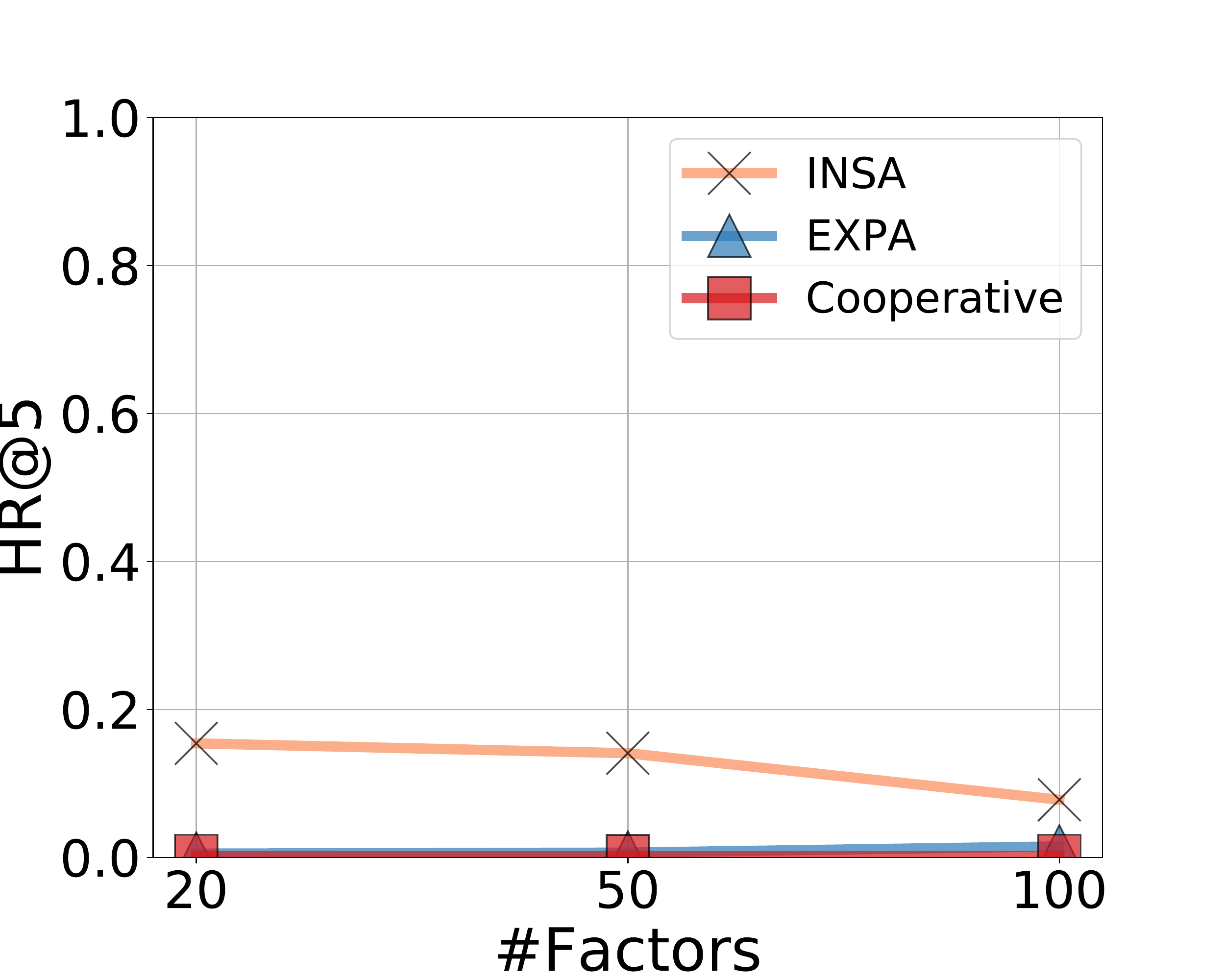}
  \caption{VBPR (Tradesy.com)}

\end{subfigure}
\caption{$\mathit{HR}@5$ of cooperative cold item and adversarial cold item by INSA/EXPA with different number of factors in VBPR and DVBPR on Amazon Men's Clothing and Tradesy.com data set.}
\label{fig:numfactor}
\vspace{-0.3cm}
\end{figure}

To study the impact of embedding length, we gradually reduce the embedding length and measure $\textit{HR}@5$.
Specifically, we conduct experiments for different numbers of factors (for VBPR in $\{20, 50, 100\}$ and for DVBPR in $\{10, 30, 50, 100\}$) with same adversarial budget (i.e., same iterations and learning rates). 
Results are presented in Figure~\ref{fig:numfactor} for VBPR and DVBPR.
We discovered that the embedding length is quite important, with evidence pointing towards systems using shorter embedding length being more vulnerable to AIP attacks.
This finding is valuable since without this knowledge a visually-aware recommender systems might use short embeddings to save storage.

To study the impact of attack strength we vary the magnitude of $\epsilon$ and measure $\textit{HR}@5$. 
We carry out experiments with $\epsilon$ in $\{4, 8, 26, 32\}$ for INSA and EXPA on subset of 100 cold item images.
Results are presented in Table~\ref{tab:eps} for all three visually-aware recommender systems.
We found that increasing $\epsilon$ from 4 to 32 leads to improved adversarial effects, but large perturbation size is not necessary in most cases for a successful attack.
Comparably, DVBPR is more sensitive to the magnitude of $\epsilon$ than the other two approaches.

Figure~\ref{fig:imageexamples} provides examples that correspond to different levels of  $\epsilon$.
This figure confirms that it is not necessary for the adversarial modifications to be highly noticeable in an image in order for an attack to be effective.

\begin{table*}[!htb]
\caption{$\mathit{HR}@5$ of a subset of 100 adversarial cold test items by INSA/EXPA with different magnitude of $\epsilon$ in DVBPR and VBPR for Amazon Men's Clothing and Tradesy.com data set. (cf. Figure~\ref{fig:imageexamples} for adversarial item images with different $\epsilon$)}
\vspace{-0.1cm}
\newcommand{\tabincell}[2]{\begin{tabular}{@{}#1@{}}#2\end{tabular}}
\renewcommand{\arraystretch}{1}
\begin{center}
\resizebox{0.8\textwidth}{!}{
\begin{tabular}{l|l!{\vrule width 1pt}ccccc|ccccc}
\toprule[1pt]
&&\multicolumn{5}{c|}{Amazon Men}&\multicolumn{5}{c}{Tradesy.com}\\
\cline{2-12}
&\textbf{Attack}&Cooperative cold&$\epsilon = 4$&$\epsilon = 8$&$\epsilon = 16$&$\epsilon = 32$&Cooperative cold&$\epsilon = 4$&$\epsilon = 8$&$\epsilon = 16$&$\epsilon = 32$\\
\midrule[1pt]
\multirow{2}{*}{\textbf{AlexRank}}
&INSA  &\multirow{2}{*}{0.0023}&0.7856 &0.8180 &0.8258 &0.8297&\multirow{2}{*}{0.0054}&0.9280 &0.9351 &0.9381 &0.9400 \\
&EXPA  &&0.0036 &0.0035 &0.0036 &0.0036&&0.0020 &0.0000 &0.0000 &0.0000\\
\hline
\multirow{2}{*}{\textbf{VBPR}}
&INSA &\multirow{2}{*}{0.0018}&0.2932 &0.3159 &0.3498 &0.3599&\multirow{2}{*}{0.0025}&0.0734 &0.0755 &0.0815 &0.0842\\
&EXPA  &&0.0008 &0.0075 &0.0135 &0.0167&&0.0005 &0.0009 &0.0009 &0.0009\\
\hline
\multirow{2}{*}{\textbf{DVBPR}}
&INSA  &\multirow{2}{*}{0.0007}&0.0049 &0.0391 &0.3706 &0.7039 &\multirow{2}{*}{0.0018}&0.0015 &0.0194 &0.1116 &0.2731\\
&EXPA  &&0.0021 &0.0088 &0.0313 &0.0554 &&0.0025 &0.0076 &0.0291 &0.0397\\
\bottomrule[1pt]
\end{tabular}
    }
\end{center}
\label{tab:eps}
\vspace{-0.3cm}
\end{table*}

\subsection{Classifier-targeted Adversarial Images}
\label{sec:taamr}

\begin{table*}[!htb]
\caption{Change in average $\mathit{HR}@5$ before and after TAaMR attack: Cold items (attack is successful if $\mathit{HR}@5$ rises) and ordinary test items (successful attacks cause a drop in $\mathit{HR}@5$) .}
\vspace{-0.1cm}
\newcommand{\tabincell}[2]{\begin{tabular}{@{}#1@{}}#2\end{tabular}}
\renewcommand{\arraystretch}{1}
\begin{center}
\resizebox{0.8\textwidth}{!}{
\begin{tabular}{l|l!{\vrule width 1pt}ccc|ccc}
\toprule[1pt]
&&\multicolumn{3}{c|}{Amazon Men}&\multicolumn{3}{c}{Tradesy.com}\\
\cline{2-8}
&&AlexRank&VBPR&DVBPR&AlexRank&VBPR&DVBPR\\
\midrule[1pt]
\multirow{2}{*}{\makecell{\textbf{Integrity} \\\textbf{dimension}}}&FGSM adversarial cold vs. cooperative cold
&$\downarrow 0.0010$&$\uparrow 0.0014$&$\uparrow 0.0070$
&$\downarrow 0.0002$&$\uparrow 0.0183$&$\downarrow 0.0008$\\
&PGD20 adversarial cold vs. cooperative cold
&$\uparrow  0.0100$&$\downarrow 0.0011$&$\downarrow 0.0005$
&$\uparrow 0.0019$&$\downarrow 0.0014$&$\downarrow 0.0004$\\
\hline
\multirow{2}{*}{\makecell{\textbf{Availability} \\\textbf{dimension}}}&FGSM ordinary test vs. ordinary test  
&$\uparrow <<0.0001$&$\downarrow <<0.0001 $&$\uparrow 0.0016$
&$\downarrow <<0.0001$&$\downarrow <<0.0001$ &$\uparrow <<0.0001$\\
&PGD20 ordinary test vs. ordinary test
&$\downarrow <<0.0001$&$\uparrow <<0.0001$&$\uparrow 0.0016$
&$\uparrow <<0.0001$&$\downarrow <<0.0001$&$\uparrow <<0.0001$\\


\bottomrule[1pt]
\end{tabular}
    }
\end{center}

\label{tab:class_hrcompare}
\vspace{-0.3cm}
\end{table*}
In Section~\ref{intro}, we pointed out that previous work on the vulnerability of recommender systems that use images has focused on classifier-targeted adversarial examples, which are already well studied in the computer vision literature.
In contrast, our EXPA and INSA approaches target the ranking mechanism, and attack the user content embedding.
In this section, we confirm that our attack poses a greater threat than the classifier-targeted attack TAaMR~\cite{DMM20}. 
The TAaMR attack works by calculating the most popular class of items using the data set on which the recommender system was trained.
As such, it can be considered an insider attack, which, like our INSA, requires knowledge that would only be available inside the company running the recommender system platform.
TAaMR also requires access to a visual feature extraction model, like our EXPA.
It uses this model as a source of class definitions and to create images that are adversarial with respect to the most popular class.
For our experiments we create class-targeted adversarial examples using two approaches from the computer vision literature,
FGSM~\cite{goodfellow2014explaining} and PGD~\cite{kurakin2016adversarial}, on pre-trained AlexNet~\cite{krizhevsky2012imagenet}.
We use the training data to calculate the most popular item classes. 
There are `jersey, T-shirt, tee shirt' (9468 interactions) for Amazon Men and `suit, suit of clothes' (19841 interactions) for Tradesy.com.
The inventory of classes is taken from ImageNet~\cite{deng2009imagenet}, as in~\cite{DMM20}.
Table~\ref{tab:class_hrcompare} shows the change in average $\textit{HR}@5$ for FGSM and PGD20 attack.
Generally, PGD20 has larger impacts on the AlexRank, and FGSM is more effective on VBPR.
Both attacks have little impact on DVBPR, because the architecture of AlexNet is fairly distinct from the CNN-F architecture~\cite{chatfield2014return} in DVBPR.
Although it uses information comparable to that used by INSA and EXPA, the adversarial impact is on par with the c-SEMA attack (cf. Table~\ref{tab:hrcompare})
This experiment shows the importance of AIP attacks, which directly attacks the user content embeddings and thereby the ranker.

\section{Defense}
\label{sec:defense}
\subsection{Adversarial Training}
\label{sec:AT}
AMR (Adversarial Multimedia Recommendation)~\cite{tang2019adversarial}
 uses the preference prediction function from VBPR (cf. Equation~\ref{eq:VBPR}) and adds on-the-fly adversarial information to the training process.
 Recall it was proposed to improve recommendation performance, but here we will study its potential for defending against AIP attacks.
The optimization of AMR is implemented by mini-batch gradient descent. 
In each step, given a subset $\mathcal{D}_{a}$ of $\mathcal{D}_{s}$ (cf. Equation~\ref{eq:tri}), AMR first perturbs the $\Theta$ to increase the loss:
\begin{equation}
\label{eq:amrperturb}
\Theta' = \underset{\Theta}{\argmax} \sum_{(u, i, j) \in \mathcal{D}_{a}} - \ln \sigma (p_{u, i} - p_{u, j})
\end{equation}
In AMR, adversarial perturbations with respect to parameters are calculated by model gradients and added to current parameters in each step.
Then it feeds forward the visual features, calculates combined loss, and back-propagates to update the parameters of the model.
Specifically, $p_{u, i}'$ and $p_{u, j}'$ are calculated by the model with perturbed parameters $\Theta'$ (cf. Equation~\ref{eq:VBPR}). Then, model parameters are updated by back-propagation based on the combined normal and adversarial loss:
\begin{equation}
\label{eq:amr}
\underset{\Theta}{\argmin} \sum_{(u, i, j) \in \mathcal{D}_{a}} - \ln \sigma (p_{u, i} - p_{u, j}) - \lambda_{\textrm{adv}} \ln \sigma (p_{u, i}' - p_{u, j}') + \lambda_{\Theta} ||\Theta||^{2}
\end{equation}
where $\lambda_{\textrm{adv}}$ is the weight hyperparameter for adversarial loss.
To train AMR, we adopt the same hyperparameters from VBPR and set $\lambda_{\text{adv}} = 1$. 

Table~\ref{tab:huge}(b) presents the $\mathit{HR}@5$ of AMR under AIP attacks and Table~\ref{tab:hrcompare}(b) presents the exact hit rate changes.
Although AMR increases the general performance by including adversarial information into training process, the $\mathit{HR}@5$ jumps noticeably when AIP attacks are applied, which means AMR is vulnerable to AIP attacks.
Our finding here is consistent with  recent research in the machine learning community~\cite{carlini2019evaluating, tramer2020adaptive}, which shows that achieving adversarial robustness is non-trivial.
\subsection{Defense by Image Compression}
\label{sec:img_compress}
In the computer vision literature, simple defenses have been shown to be effective against adversarial images that cause neural classifiers to misclassify~\cite{guo2018countering, das2018shield, dziugaite2016study, xu2017feature}.
Here, we evaluated two common defenses: JPEG compression and bit depth reduction in order to test whether they are effective against AIP attacks.
These are known to be able to erase the effect of image perturbations.
We carried out an evaluation by applying progressively stronger versions of the defense to adversarial images.
We used a 100-item subset of our larger test set.
We do not evaluate SEMA, since the semantic attack does not involve perturbations and if these defenses would destroy the effectiveness of SEMA they would destroy the usefulness of all images to the recommender system.

In Table~\ref{tab:defense}, we visualize the level of strength of defense that must be applied to the adversarial image in order for its rank to be lowered to the average $\mathit{HR}@5$ of a cooperative image.
If the defenses presented effective protection against adversarial item images 
then we would expect the $\blacktriangle$ and $\bigstar$ to appear consistently to the far left in the boxes.
This is clearly not the case.
We see in Table~\ref{tab:defense} that INSA is more difficult to defeat than EXPA, which is expected because it leverages insider knowledge.
However, EXPA is clearly not easy to beat across the board.
It is important to note that this test is a strong one.
If these defenses would be applied in practice, they would need to be applied to all images and not just adversarial images.
Image content becomes indistinguishable as compression increases, and an image 10\% the size of the original image or encoded with only 2-3 bits can be expected to contain little to no item information.

\begin{table}[t]
\caption{Visualization of the level at which a defense is successful at lowering the $\mathit{HR}@5$ of an adversarial cold  item equal or less the average $\mathit{HR}@5$ of a cooperative cold  item. For JPEG compression, levels are specified as compression percents and for bit depth reduction levels are specified as number of bits with which the image is encoded. ($\blacktriangle$: Amazon.com; $\bigstar$: Tradesy.com)}
\newcommand{\tabincell}[2]{\begin{tabular}{@{}#1@{}}#2\end{tabular}}
\renewcommand{\arraystretch}{1}
\begin{center}
\resizebox{0.48\textwidth}{!}{
\begin{tabular}{ll|ccccc|cccccc}
\toprule[1pt]
&\multicolumn{1}{c|}{}&\multicolumn{5}{c|}{JPEG compression}&\multicolumn{6}{c}{Bit depth reduction}\\
\cline{3-13}
&&90&70&50&30&10&7&6&5&4&3&2\\
\midrule[1pt]
\multirow{2}{*}{AlexRank}
&INSA&&&&&&&&&&&\\
&EXPA&&&&&&&&&&$\blacktriangle$&\\
\hline
\multirow{2}{*}{VBPR}
&INSA&&&&&&&&&&&\\
&EXPA&$\blacktriangle \bigstar$&&&&&&&&&$\blacktriangle \bigstar$&\\
\hline
\multirow{2}{*}{DVBPR} 
&INSA&&&&&&$\bigstar$&&&&&\\
&EXPA&&&&&$\blacktriangle$&&&&&&$\blacktriangle \bigstar$\\

\bottomrule[1pt]
\end{tabular}
    }
\end{center}
\label{tab:defense}
\vspace{-0.5cm}
\end{table}

\section{Conclusion and Outlook}
This paper has investigated the vulnerability at the core of Top-N recommenders that use images to address cold start.
We have shown that Adversarial Item Promotion (AIP) attacks allow unscrupulous merchants to artificially raise the rank of their products when a visually-aware recommender system is used for candidate ranking.
Our investigation has led us to conclude that AIP attacks are a potential threat with clear practical implications.
Compared with existing profile injection attacks~\cite{o2002promoting, lam2004shilling, mobasher2007toward, burke2015robust} and poisoning attacks~\cite{li2016data, fang2018poisoning, christakopoulou2019adversarial, fang2020influence, tang2019adversarial} that promote items by injecting fake profiles, AIP only needs to modify the descriptive image of the item. 
Effective AIP attacks are easy to mount, as demonstrated by the minimal scale attack that we have studied here (cf. attack model in Section~\ref{sec:attackmodel}).
In short, our work reveals that the promise of hybrid recommender systems to provide a higher degree of robustness~\cite{mobasher2007toward} is not an absolute, and that we must proceed with caution when using images to address cold start.

Future work should dive more deeply into connection between adversarial items and user experience with the recommender system.
One aspect is the relevance of adversarial items to users.
Like any cold start item, users click an adversarial cold start item because it piques their interest.
As an adversarial item accumulates more clicks, and enters more users' personalized lists, the main issue may be not be relevance, but rather fair competition with other potentially relevant items.

Another aspect related to user experience is the impact of image quality.
If users have sensitivities that cause them to avoid products with images affected by perturbations, then attackers would need to back off to weaker attacks that make perturbations unnoticeable.
In this case, defenses such as adversarial training could be more effective.
More work is needed to understand approaches such as SEMA, which do not involve trading off image quality and attack strength.
Alternatively, approaches that make adversarial images effective yet non-suspicious, such as~\cite{zhao2019towards, joshi2019semantic}, can also be studied.

Future work must develop effective defenses against AIP attacks. 
An approach that easily comes to mind is the use a gatekeeper classifier to flag adversarial images at the moment that merchants upload them.
It is clear that for SEMA such a classifier would be difficult to build, since SEMA attacks are created in a natural manner and are indistinguishable from cooperative images.
For INSA and EXPA, a gateway filter could be built if the exact specifications of the adversarial attack, including the parameter settings, were known.
However, we need to be aware that in the worst case scenario where the information of the gatekeeper is available (i.e., white-box scenario), variants on INSA and EXPA can still bypass such a classifier by constructing new loss functions~\cite{carlini2017adversarial}.

We have shown in our paper (cf. Section~\ref{sec:AT}) that simply incorporating on-the-fly adversarial information into model training cannot guarantee a robust recommender.
In addition, adversarial training requires strict hypothesis about the attack strength ($\epsilon$)~\cite{madry2017towards}, and it also needs vast computational resources in practice~\cite{xie2019intriguing}.
Therefore, building a robust visually-aware recommender system is non-trivial and needs more research attention.

Future work must look at the impact of multipliers.
If a single item has multiple descriptive images, attacks are more likely to go unnoticed, in particular semantic attacks that require no perturbations.
Further, multiple merchants (or fake merchant profiles) could collaborate in a collusion attack.

Finally, we note that although, here, we have focused on e-commerce, entertainment recommender systems are vulnerable: an adversarial signal could be embedded into a thumbnail or the content itself. 
In sum, AIP attacks constitute an important, practical risk of using images in recommender systems and serious challenges remain to be addressed.

\bibliographystyle{ACM-Reference-Format}
\bibliography{references}

\end{document}